\documentclass[aps,prx,amsmath,amssymb,superscriptaddress,twocolumn,longbibliography]{revtex4-2}
\usepackage{bbm}
\usepackage{graphicx}
\usepackage{dcolumn}
\usepackage{bm}
\usepackage{subfigure}
\usepackage{amsmath}
\usepackage{stmaryrd}
\usepackage{times}
\usepackage{graphicx}
\usepackage{dcolumn}
\usepackage{color}
\usepackage{makecell}
\usepackage{tabularx,multirow}
\usepackage{appendix}
\usepackage{amssymb}
\usepackage{pifont}
\usepackage[colorlinks,linkcolor=blue,citecolor=blue,urlcolor=black]{hyperref}
\begin{document}
\title{Spin Conductivity: Insight From Perturbation Theory}
\author{Xu Chen}
\affiliation{School of Physics, Harbin Institute of Technology, Harbin 150001, China}
\author{Mingbo Dou}
\affiliation{School of Physics, Harbin Institute of Technology, Harbin 150001, China}
\author{Yu Sui}
\affiliation{School of Physics, Harbin Institute of Technology, Harbin 150001, China}
\author{Xianjie Wang}
\affiliation{School of Physics, Harbin Institute of Technology, Harbin 150001, China}
\affiliation{Heilongjiang Provincial Key Laboratory of Advanced Quantum Functional Materials and Sensor Devices, Harbin 150001, China}
\author{L. L. Tao}
\email{Contact author: lltao@hit.edu.cn}
\affiliation{School of Physics, Harbin Institute of Technology, Harbin 150001, China}
\affiliation{Heilongjiang Provincial Key Laboratory of Advanced Quantum Functional Materials and Sensor Devices, Harbin 150001, China}
\date{\today}
\begin{abstract}
The investigation on the spin conductivity has now been blooming in burgeoning altermagnetic materials. Here, an alternative and elementary derivation for the spin conductivity expression is presented. This is done within the context of perturbation theory. It indicates that our derived expression for the $\mathcal{T}$-even ($\mathcal{T}$ for time reversal) spin conductivity agrees exactly with those derived from the Kubo linear response theory and quantum kinetic theory. In addition, the expression for the $\mathcal{T}$-odd spin conductivity in the clean limit is exactly consistent with those presented previously. We show that the $\mathcal{T}$-odd ($\mathcal{T}$-even) spin conductivity arises from the contribution of zeroth-order (first-order) wave function correction. Our work provides some physical insight into the spin conductivity from the perturbation theory and shines more light on the charge-to-spin conversion mechanism in magnetic materials.
\end{abstract}
\maketitle
\section{Introduction}
The charge-to-spin conversion enables all-electric generation of spin currents and plays a very important role in spintronics\cite{rmp1213,npj27,nrm258}. For example, the spin Hall effect (SHE) characterized by a transverse spin current in response to a longitudinal charge current is essential for many intriguing physical phenomena, such as the current-induced spin polarization\cite{jpcm179,prb085438}, the spin–orbit torque\cite{prb212405,prb094422,prb094424,nature189,science555,prl096602,rmp035004}, the spin Hall magnetoresistance effect\cite{prb144411,np570}, etc. In particular, the spin-orbit torque exerted by a spin current enable the current-induced magnetization switching beneficial for high-speed and low-power spintronic devices\cite{apl120502,pms100761,npj8,apr011305}. However, as enforced by the orthonormality relations between the charge current, the spin current and its spin polarization, the spin current generated by the conventional SHE typically reveals the in-plane spin polarization, which is not favorable to switch the perpendicular magnetization\cite{apl120502,pms100761,npj8,apr011305}. Moreover, the charge-to-spin conversion efficiency is dominated by the strength of the spin-orbit coupling (SOC), which is small compared to the magnetic exchange coupling. In this context, exploration of novel charge-to-spin conversion mechanism is desirable.

To first order in an applied electric field $\bm{\mathcal{E}}$, the spin current density $\mathbf{J}_s$ can be expressed as $J_{b}^a=\sigma_{bc}^a\mathcal{E}_c$\cite{prb024410}, where the spin conductivity $\sigma_{bc}^a$ is a third-rank tensor and describes that a charge current along the $c$ direction induces a spin current along the $b$ direction with the spin polarization along the $a$ direction. The indices $a, b, c=x, y, z$ denote Cartesian components and a summation over repeated indices is understood. It is established that $\sigma_{bc}^{a}$ can be classified into two categories according to whether $\sigma_{bc}^{a}$ is $\mathcal{T}$-odd or $\mathcal{T}$-even under the time reversal $\mathcal{T}$, that is, $\sigma_{bc}^{a}=\sigma_{bc}^{a,\text{odd}}+\sigma_{bc}^{a,\text{even}}$\cite{prb174423,prb134402,prl187204,na627,prl127701,prb224401,npj46}. $\sigma_{bc}^{a,\text{even}}$ represents the conventional SHE commonly investigated in nonmagnetic materials while $\sigma_{bc}^{a,\text{odd}}$ corresponds to the magnetic SHE absent in nonmagnetic materials. Another major difference between $\sigma_{bc}^{a,\text{odd}}$ and $\sigma_{bc}^{a,\text{even}}$ is that the SOC is essential for $\sigma_{bc}^{a,\text{even}}$. The magnetic SHE was theoretically studied in the non-collinear antiferromagnets Mn$_3$Sn and Mn$_3$Ir\cite{prl187204} and was observed experimentally in Mn$_3$Sn\cite{na627}, which shows that the magnetic SHE is reversed upon the triangularly ordered moments switching. Recently, the investigation of the magnetic SHE has been boosting in altermagnets\cite{prx031042,prx040501,afm2409327,jcas2257,nrm2025,nature837}. As compared to the conventional ferromagnetism and antiferromagnetism, the altermagnetism emerges as a distinct magnetic phase combining the compensated magnetization with sizable nonrelativistic spin splitting (of the order of $1.0$ eV)\cite{jpsj123702,prb014422,nc2846}. In particular, a distinct spin-splitter torque (SST)\cite{prl127701} effect was theoretically proposed based on the spin-splitter effect in the altermagnet RuO$_2$ and the magnetization switching driven by the SST has been observed experimentally in the RuO$_2$/Py structure\cite{prl197202,prl137201,nc5646,nc1309}.

The formulas to calculate the spin conductivity were derived mostly from the Kubo linear response theory\cite{prb174423,prb014416} or quantum kinetic theory\cite{na627}. In this work, we propose an alternative but simplified approach to derive the spin conductivity expression within the framework of perturbation theory\cite{Vanderbilt}. Some discussion are devoted to the comparison with those presented previously and the nature of distinct contributions, which helps in clarifying the essential physics of the charge-to-spin conversion mechanism.

\section{Adiabatic perturbation theory\label{sec2}}
The starting point is the following elementary Hamiltonian for a generic periodic system with an applied uniform electric field $\bm{\mathcal{E}}$\cite{Vanderbilt,ase14818}:
\begin{equation}\label{eq-1}
  \mathcal{H}=\mathcal{H}_0+e\bm{\mathcal{E}}\cdot\mathbf{r},
\end{equation}
where $\mathcal{H}_0$ is the Hamiltonian for a periodic system and the second term $e\bm{\mathcal{E}}\cdot\mathbf{r}$ given in terms of the electron charge $e$ and the position operator $\mathbf{r}$ is the electric potential operator induced by $\bm{\mathcal{E}}$. In the crystal momentum representation, the eigenstates of $\mathcal{H}_0$ are Bloch functions $\psi_{n\mathbf{k}}(\mathbf{r})$ denoted by the band index $n$ and the wave vector $\mathbf{k}$. Application of $\bm{\mathcal{E}}$, $\mathbf{k}$ is time $t$ dependent according to the equation of motion $\hbar\dot{\mathbf{k}}=-e\bm{\mathcal{E}}$, where $\hbar$ is the reduced Planck's constant and a dot over $\mathbf{k}$ indicates the first derivative with respect to $t$. This implies that $\mathcal{H}(t)$ is the functional of $\mathbf{k}(t)$, which we denote as $\mathcal{H}(\mathbf{k}(t))$. It is assumed that $\mathcal{H}(t)$ changes slowly with $t$, keeping the system in its instantaneous eigenstate, which is known as the adiabatic approximation. Thus, we have the Schrödinger equation $\mathcal{H}(t)|\Psi(t)\rangle=\epsilon(t)|\Psi(t)\rangle$, where $\epsilon(t)$ ($|\Psi(t)\rangle$) is the instantaneous eigenvalue (eigenstate). At $t=0$, it is assumed that $|\Psi(t=0)\rangle=|\psi_{n\mathbf{k}}\rangle$ and $\mathcal{H}(t=0)|\psi_{n\mathbf{k}}\rangle=\epsilon_{n\mathbf{k}}|\psi_{n\mathbf{k}}\rangle$. In the spirit of the adiabatic perturbation theory and to first order in $\dot{k}_a$, $|\Psi(t)\rangle$ takes the form\cite{Vanderbilt}
\begin{equation}\label{eq-2}
  |\Psi(t)\rangle=e^{i[\theta_{n\mathbf{k}}(t)+\gamma_{n\mathbf{k}}(t)]}(|\psi_{n\mathbf{k}}\rangle+\dot{k}_a|\delta\psi_{nk_a}\rangle),
\end{equation}
where $\theta_{n\mathbf{k}}(t)=-\frac{1}{\hbar}\int^t_0\epsilon_n(\mathbf{k}(t'))dt'$ is the dynamical phase while $\gamma_{n\mathbf{k}}(t)=i\int^t_0\langle\psi_n(\mathbf{k}(t'))|\dot{\psi_n}(\mathbf{k}(t'))\rangle dt'=i\int^{\mathbf{k}(t)}_{\mathbf{k}(0)}\mathbf{A}_n(\mathbf{k'})\cdot d\mathbf{k}'$ is the geometric phase or the Berry phase\cite{Berry,rmd1959} and $A_{nk_a}\equiv i\langle\psi_{n\mathbf{k}}|\partial_{k_a}\psi_{n\mathbf{k}}\rangle$ is the Berry connection. The index $a = x, y, z$ denote Cartesian components and a summation over repeated indices is understood. $|\delta \psi_{nk_a}\rangle$ is to be determined. We substitute Eq. (\ref{eq-2}) into the time-dependent Schrödinger equation $i\hbar\partial_t|\Psi(t)\rangle=\mathcal{H}|\Psi(t)\rangle$. Then, we obtain (see Appendix \ref{secA} for details)
\begin{equation}\label{eq-3}
  (\epsilon_{n\mathbf{k}}-\mathcal{H})|\delta\psi_{nk_a}\rangle=-i\hbar(\partial_{k_a}+iA_{nk_a})|\psi_{n\mathbf{k}}\rangle,
\end{equation}
The solution of Eq. (\ref{eq-3}) is (see Appendix \ref{secB} for details)
\begin{equation}\label{eq-4}
  |\delta\psi_{nk_a}\rangle=-i\hbar\sum_{m\neq n}\frac{|\psi_{m\mathbf{k}}\rangle\langle\psi_{m\mathbf{k}}|}{\epsilon_{n\mathbf{k}}-\epsilon_{m\mathbf{k}}}|\partial_{k_a}\psi_{n\mathbf{k}}\rangle.
\end{equation}
This expression is to be substituted into Eq. (\ref{eq-2}). We finally have
\begin{equation}\label{eq-5}
  |\Psi\rangle=e^{i(\theta_{n\mathbf{k}}+\gamma_{n\mathbf{k}})}(|\psi_{n\mathbf{k}}\rangle-i\hbar\dot{k}_aT_{n\mathbf{k}}|\partial_{k_a}\psi_{n\mathbf{k}}\rangle),
\end{equation}
and the operator $T_n$ is defined as
\begin{equation}\label{eq-6}
  T_{n\mathbf{k}}\equiv\sum_{m\neq n}\frac{|\psi_{m\mathbf{k}}\rangle\langle\psi_{m\mathbf{k}}|}{\epsilon_{n\mathbf{k}}-\epsilon_{m\mathbf{k}}}.
\end{equation}
We now proceed to calculate the expectation value of the spin current operator $\mathcal{J}^a_b=\hbar\{\sigma_a, v_b\}/4$ given in terms of the spin (velocity) operator $\sigma_a$ ($v_b$). Again, the second-order terms proportional to $\dot{k}_a\dot{k}_b$ will be ignored. This can be expressed as
\begin{equation}\label{eq-7}
\begin{aligned}
  \langle\mathcal{J}^a_b\rangle_{n\mathbf{k}}&=\langle\psi_{n\mathbf{k}}|\mathcal{J}^a_b|\psi_{n\mathbf{k}}\rangle-i\hbar\dot{k}_c[\langle\psi_{n\mathbf{k}}|\mathcal{J}^a_bT_{n\mathbf{k}}|\partial_{k_c}\psi_{n\mathbf{k}}\rangle\\
  &-\langle\partial_{k_c}\psi_{n\mathbf{k}}|T_{n\mathbf{k}}\mathcal{J}^a_b|\psi_{n\mathbf{k}}\rangle]\\
  &=\langle\psi_{n\mathbf{k}}|\mathcal{J}^a_b|\psi_{n\mathbf{k}}\rangle+2\hbar\dot{k}_c\text{Im}\langle\psi_{n\mathbf{k}}|\mathcal{J}^a_bT_{n\mathbf{k}}|\partial_{k_c}\psi_{n\mathbf{k}}\rangle\\
  &=\langle\psi_{n\mathbf{k}}|\mathcal{J}^a_b|\psi_{n\mathbf{k}}\rangle\\
  &-2e\mathcal{E}_c\sum_{m\neq n}\frac{\text{Im}[\langle\psi_{n\mathbf{k}}|\mathcal{J}^a_b|\psi_{m\mathbf{k}}\rangle\langle\psi_{m\mathbf{k}}|\partial_{k_c}\mathcal{H}|\psi_{n\mathbf{k}}\rangle]}{(\epsilon_{n\mathbf{k}}-\epsilon_{m\mathbf{k}})^2}.
\end{aligned}
\end{equation}
In the last step, we have used the relation
\begin{equation}\label{eq-8}
\langle\psi_{m\mathbf{k}}|\partial_{k_c}\psi_{n\mathbf{k}}\rangle=\frac{\langle\psi_{m\mathbf{k}}|\partial_{k_c}\mathcal{H}|\psi_{n\mathbf{k}}\rangle}{\epsilon_{n\mathbf{k}}-\epsilon_{m\mathbf{k}}},
\end{equation}
and its proof is presented in Appendix \ref{secC}. Then, the spin current density $J^a_b$ can be obtained with the use of the Boltzmann transport theory. To first order in $\mathcal{E}$, the distribution function $f$ can be expanded as $f=f_0+f_1$, where $f_0$ and $f_1$ are the equilibrium (Fermi distribution function) and first-order distribution functions, respectively. Here we consider the relaxation-time $\tau$ approximation and $f_1$ is given by
\begin{equation}\label{eq-9}
  f_1=\frac{e\tau}{\hbar}\mathcal{E}_c\frac{\partial f_0}{\partial k_c}=e\tau\mathcal{E}_c\langle\psi_{n\mathbf{k}}|v_c|\psi_{n\mathbf{k}}\rangle\frac{\partial f_0}{\partial\epsilon_{n\mathbf{k}}}.
\end{equation}
It is to be noted that $f_0(\epsilon_{n\mathbf{k}}, \epsilon_{F})$ is the function of $\epsilon_{n\mathbf{k}}$ and the Fermi energy $\epsilon_{F}$. Then, $J^a_b$ is summed over occupied states by combining Eqs. (\ref{eq-7}) and (\ref{eq-9})
\begin{equation}\label{eq-10}
\begin{aligned}
  J^a_b&=\sum_{n\mathbf{k}}\langle\mathcal{J}^a_b\rangle_{n\mathbf{k}}f=\sum_{n\mathbf{k}}\langle\psi_{n\mathbf{k}}|\mathcal{J}^a_b|\psi_{n\mathbf{k}}\rangle f_1\\
  &-2e\mathcal{E}_c\sum_{mn\mathbf{k}}^{m\neq n}\frac{\text{Im}[\langle\psi_{n\mathbf{k}}|\mathcal{J}^a_b|\psi_{m\mathbf{k}}\rangle\langle\psi_{m\mathbf{k}}|\partial_{k_c}\mathcal{H}|\psi_{n\mathbf{k}}\rangle]}{(\epsilon_{n\mathbf{k}}-\epsilon_{m\mathbf{k}})^2}f_0\\
  &=e\tau\mathcal{E}_c\sum_{n\mathbf{k}}\langle\psi_{n\mathbf{k}}|\mathcal{J}^a_b|\psi_{n\mathbf{k}}\rangle\langle\psi_{n\mathbf{k}}|v_c|\psi_{n\mathbf{k}}\rangle\frac{\partial f_0}{\partial\epsilon_{n\mathbf{k}}}\\
  &-2e\hbar\mathcal{E}_c\sum_{mn\mathbf{k}}^{m\neq n}\frac{\text{Im}[\langle\psi_{n\mathbf{k}}|\mathcal{J}^a_b|\psi_{m\mathbf{k}}\rangle\langle\psi_{m\mathbf{k}}|v_c|\psi_{n\mathbf{k}}\rangle]}{(\epsilon_{n\mathbf{k}}-\epsilon_{m\mathbf{k}})^2}f_0.
\end{aligned}
\end{equation}
With the definition of $\sigma_{bc}^a$ from $J_{b}^a=\sigma_{bc}^a\mathcal{E}_c$, we have
\begin{equation}\label{eq-11}
\begin{aligned}
  \sigma_{bc}^a=&e\tau\sum_{n\mathbf{k}}\langle\psi_{n\mathbf{k}}|\mathcal{J}^a_b|\psi_{n\mathbf{k}}\rangle\langle\psi_{n\mathbf{k}}|v_c|\psi_{n\mathbf{k}}\rangle\frac{\partial f_0}{\partial\epsilon_{n\mathbf{k}}}\\
  &-2e\hbar\sum_{mn\mathbf{k}}^{m\neq n}\frac{\text{Im}[\langle\psi_{n\mathbf{k}}|\mathcal{J}^a_b|\psi_{m\mathbf{k}}\rangle\langle\psi_{m\mathbf{k}}|v_c|\psi_{n\mathbf{k}}\rangle]}{(\epsilon_{n\mathbf{k}}-\epsilon_{m\mathbf{k}})^2}f_0.
\end{aligned}
\end{equation}
It is seen that $\sigma_{bc}^a$ is contributed by two distinct terms. Under the time reversal $\mathcal{T}$, since $v_c\rightarrow-v_c$ and $\mathcal{J}^a_b\rightarrow\mathcal{J}^a_b$, the first term is odd under $\mathcal{T}$ and shall be denoted as $\sigma_{bc}^{a,\text{odd}}$. We denote the numerator in the second term as $\text{Im}[Z]$, where the complex number $Z$ is proportional to $\mathcal{J}^a_bv_c$. Thus, we have $\text{Im}[Z]\rightarrow\text{Im}[-Z^*]=\text{Im}[Z]$ under $\mathcal{T}$. This indicates that the second term is even under $\mathcal{T}$ and shall be denoted as $\sigma_{bc}^{a,\text{even}}$. Then, we have
\begin{equation}\label{eq-12}
  \sigma_{bc}^{a,\text{odd}}=e\tau\sum_{n\mathbf{k}}\langle\psi_{n\mathbf{k}}|\mathcal{J}^a_b|\psi_{n\mathbf{k}}\rangle\langle\psi_{n\mathbf{k}}|v_c|\psi_{n\mathbf{k}}\rangle\frac{\partial f_0}{\partial\epsilon_{n\mathbf{k}}},
\end{equation}
and
\begin{equation}\label{eq-13}
  \sigma_{bc}^{a,\text{even}}=-2e\hbar\sum_{mn\mathbf{k}}^{m\neq n}\frac{\text{Im}[\langle\psi_{n\mathbf{k}}|\mathcal{J}^a_b|\psi_{m\mathbf{k}}\rangle\langle\psi_{m\mathbf{k}}|v_c|\psi_{n\mathbf{k}}\rangle]}{(\epsilon_{n\mathbf{k}}-\epsilon_{m\mathbf{k}})^2}f_0.
\end{equation}
From Eq. (\ref{eq-7}), it is unveiled that the $\mathcal{T}$-odd spin conductivity arises from the contribution of the zeroth-order wave function while the $\mathcal{T}$-even spin conductivity is dominated by the contribution from the first-order wave function correction. In the nonrelativistic limit, an electron is described by a two-component wave function and the spin current $\mathbf{J}_s$ can be written in terms of the charge current as $J_{b}^a=-\hbar/(2e)(J_b^{\uparrow}-J_b^{\downarrow})=\sigma_{bc}^{a,\text{odd}}\mathcal{E}_c$\cite{prb024410}, where $J_b^{\uparrow}$ and $J_b^{\downarrow}$ correspond to the ``up" spin ($\uparrow$) and ``down" spin ($\downarrow$) charge currents, respectively. This indicates that $\sigma_{bc}^{a,\text{odd}}$ is finite provided that the charge current is spin polarized, that is, the conductivity is spin polarized. We conclude that $\sigma_{bc}^{a,\text{odd}}$ in the nonrelativistic limit could be finite in magnetic materials with spin-split bands.

It is instructive to compare the present results Eqs. (\ref{eq-12}) and (\ref{eq-13}) with those obtained previously\cite{prb174423,prb134402,prl187204,na627,prl127701,prb224401,npj46}. Note that the $\mathcal{T}$-even contribution $\sigma_{bc}^{a,\text{even}}$ known as the spin Hall conductivity has exactly the same form with previous work\cite{prb174423,prb134402,prl187204,na627,prl127701,prb224401,npj46}. Thus, we shall focus here on $\sigma_{bc}^{a,\text{odd}}$. From Eq. (\ref{eq-12}), $\sigma_{bc}^{a,\text{odd}}$ contains only intraband matrix elements, which is contrary to that $\sigma_{bc}^{a,\text{even}}$ has both intraband and interband contributions. It is worth noting here that our derived Eq. (\ref{eq-12}) for $\sigma_{bc}^{a,\text{odd}}$ is exactly the same as the expression given in Ref. \cite{na627}, which is based on the quantum kinetic theory and relaxation time $\tau$ approximation. Based on the Kubo linear response theory and Green's function approach, $\sigma_{bc}^{a,\text{odd}}$ was presented previously as\cite{prb174423,prb134402,prl187204}
\begin{equation}\label{eq-14}
    \sigma_{bc}^{a,\text{odd}} = -\frac{e\hbar}{\pi}\sum_{mn\mathbf{k}}\frac{\text{Re}[\langle\psi_{n\mathbf{k}}|\mathcal{J}^a_b|\psi_{m\mathbf{k}}\rangle\langle\psi_{m\mathbf{k}}|v_c|\psi_{n\mathbf{k}}\rangle]\Gamma^2}{[(\epsilon_F-\epsilon_{n\mathbf{k}})^2+\Gamma^2][(\epsilon_F-\epsilon_{m\mathbf{k}})^2+\Gamma^2]},
\end{equation}
where $\Gamma=\hbar/(2\tau)$ is the broadening parameter. This suggests that $\sigma_{bc}^{a,\text{odd}}$ contains both intraband and interband contributions. However, in the clean limit of $\Gamma\to0$, with the aid of the identity $\lim\limits_{\Gamma\to0}\frac{\Gamma}{\pi}\frac{1}{(\epsilon_F-\epsilon_{n\mathbf{k}})^2+\Gamma^2}=\delta(\epsilon_F-\epsilon_{n\mathbf{k}})$, Eq. (\ref{eq-14}) is reduced to\cite{prb174423}
\begin{equation}\label{eq-15}
    \sigma_{bc}^{a,\text{odd}} = -e\tau\sum_{n\mathbf{k}}\langle\psi_{n\mathbf{k}}|\mathcal{J}^a_b|\psi_{n\mathbf{k}}\rangle\langle\psi_{n\mathbf{k}}|v_c|\psi_{n\mathbf{k}}\rangle\delta(\epsilon_F-\epsilon_{n\mathbf{k}}),
\end{equation}
which agrees exactly with Eq. (\ref{eq-12}) in the zero-temperature limit by noting that $\partial f_0/\partial\epsilon_{n\mathbf{k}}=-\delta(\epsilon_F-\epsilon_{n\mathbf{k}})$.

On the other hand, it has to be stressed that the vertex correction to capture the side jump and the skew scattering\cite{prb014416} is omitted by both previous\cite{prb174423,prb134402,prl187204,na627,prl127701,prb224401,npj46} and our work. This is due to the fact that the Green’s function formalism approximates the effect of disorder by a constant band broadening $\Gamma$\cite{prb174423} while the quantum kinetic formalism\cite{na627} or the adiabatic perturbation theory adopts the constant relaxation time approximation. The essential vertex correction may have a significant effect on the spin conductivity, which calls for further exploration.

\section{Nonrelativistic spin conductivity\label{sec3}}
\begin{figure}
\includegraphics[width=0.45\textwidth]{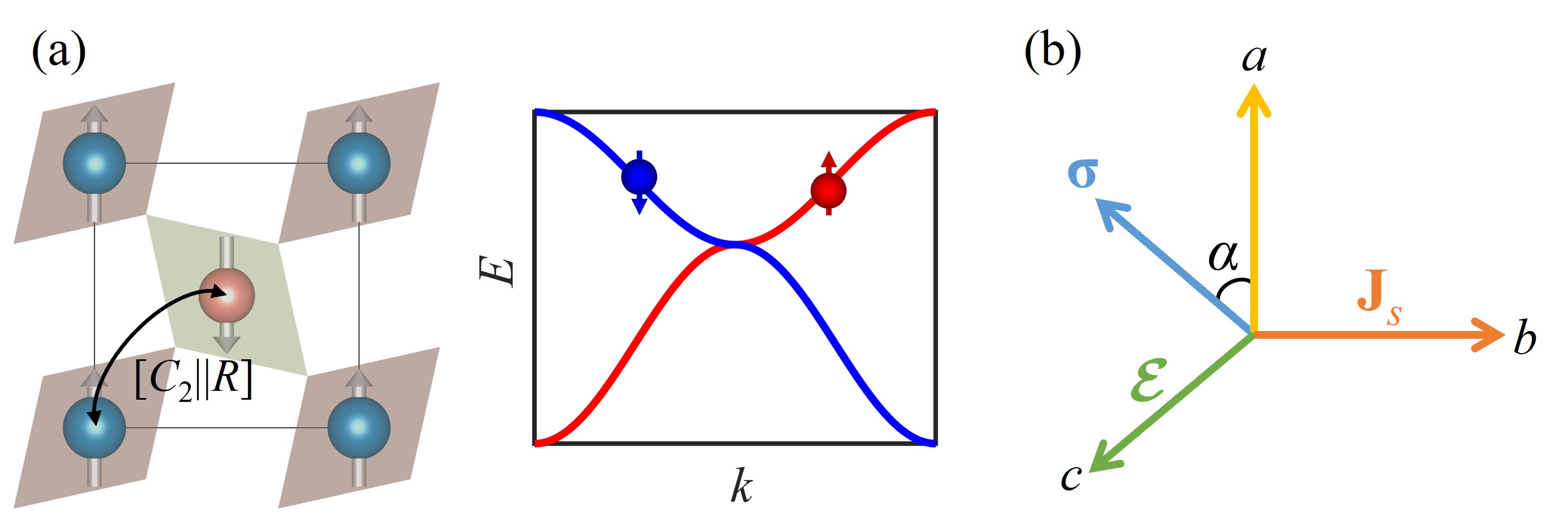}%
\caption{\label{f-1} (a) Schematic magnetic structure and spin-split bands of an altermagnet. The two opposite-spin sublattices are connected by the spin-group operation $[C_2||R]$, where $C_2$ and $R$ represent the spin-space and real-space operations (rotation or mirror), respectively. Red and blue lines indicate two spin channels. (b) Schematic orientations of an electric field $\bm{\mathcal{E}}$ along the $c$ direction, a spin current $\mathbf{J}_s$ along the $b$ direction and the spin polarization $\bm{\sigma}$ in Cartesian coordinates. $\alpha$ indicates the angle between the $a$-axis and $\bm{\sigma}$ (spin quantization axis).}
\end{figure}

We now turn to consider the spin conductivity in the nonrelativistic limit. Given the fact that the $\mathcal{T}$-even spin conductivity is vanishing without SOC and the $\mathcal{T}$ symmetry must be broken for the $\mathcal{T}$-odd spin conductivity, it is meaningful to explore the $\mathcal{T}$-odd spin conductivity in magnetic materials. In particular, the sizable spin splitting can be induced by the real-space coordination environments in altermagnets in the absence of SOC [see Fig. \ref{f-1}(a)]\cite{prx031042,prx040501}. It is enlightening to examine the $\mathcal{T}$-odd spin conductivity in the nonrelativistic limit for altermagnets. For altermagnets without SOC, we attach a spin index $s$ ($s=\uparrow, \downarrow$) to the Bloch function $|\psi^s_{n\mathbf{k}}\rangle$. Assume that the angle between the $a$-axis and the spin polarization $\bm{\sigma}$ is $\alpha$ [see Fig. \ref{f-1}(b)]. Then, we have\cite{prb224423,apr031412}
\begin{equation}\label{eq-16}
    \sigma_{bc}^{a,\text{odd}} = -\frac{\hbar}{2e}(\bar{\sigma}_{bc}^{\uparrow}-\bar{\sigma}_{bc}^{\downarrow})\cos\alpha,
\end{equation}
where $\bar{\sigma}_{bc}^s$ is the spin-resolved conductivity.

\begin{figure*}
\includegraphics[width=0.9\textwidth]{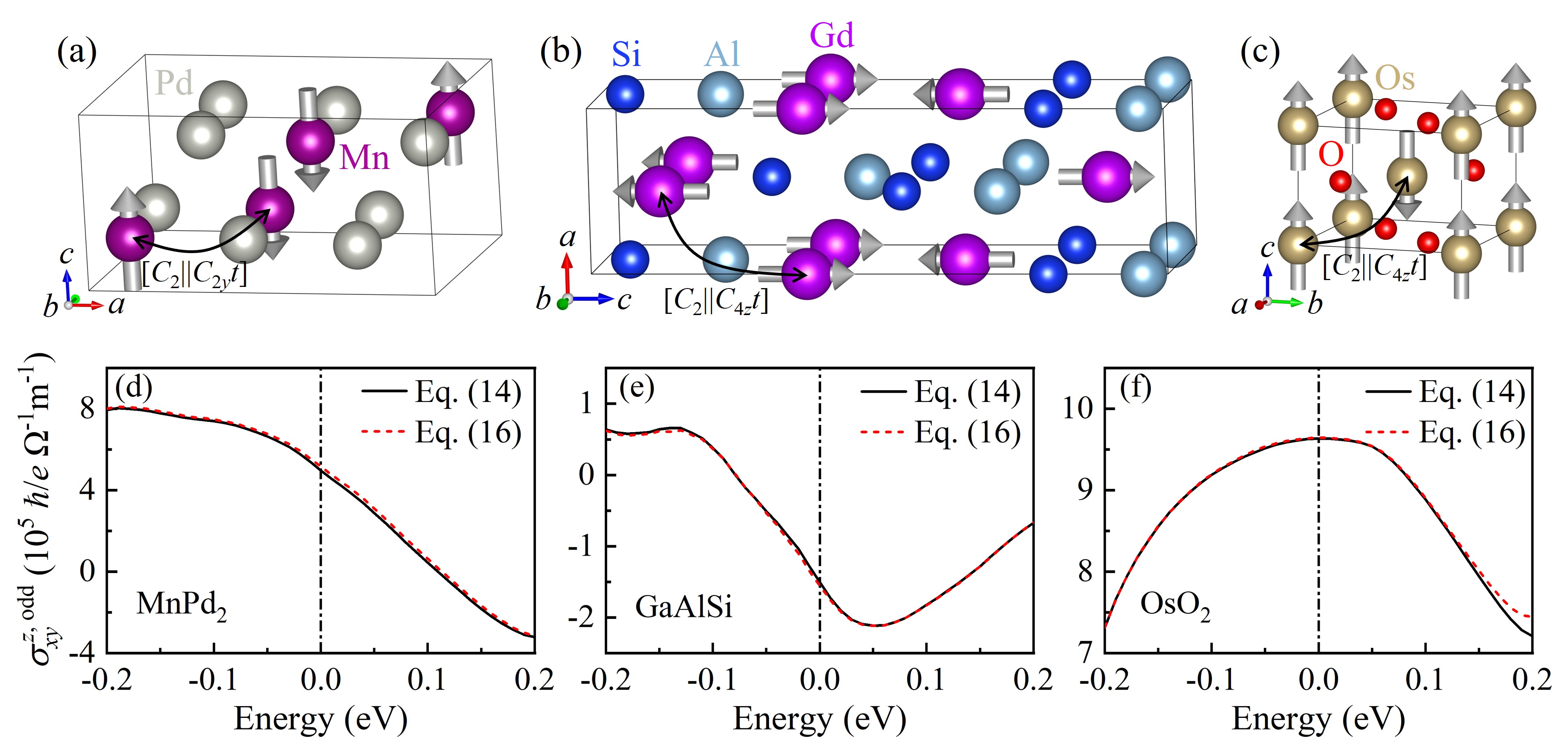}%
\caption{\label{f-2} Crystal structures for the orthorhombic MnPd$_2$ (a), tetragonal GdAlSi (b), and tetragonal OsO$_2$ (c). The gray arrows denote the magnetic moments and the opposite-spin sublattices are connected by the spin-group operations $[C_2||C_{2y}t]$ with the translation $t=(1/2, 1/2, 1/2)$ for MnPd$_2$, $[C_2||C_{4z}t]$ with $t=(0, 1/2, 1/4)$ for GdAlSi and $[C_2||C_{4z}t]$ with $t=(1/2, 1/2, 1/2)$ for OsO$_2$. $\sigma_{xy}^{z, \text{odd}}$ without SOC as a function of energy for MnPd$_2$ (d), GdAlSi (e) and OsO$_2$ (f). Solid lines were obtained using Eq. (\ref{eq-14}) with $\Gamma=10$ meV while dashed lines were obtained using Eq. (\ref{eq-16}) with $\tau=\hbar/(2\Gamma)\approx33$ fs (corresponding to $\Gamma=10$ meV) and a temperature of $50$ K in the Fermi distribution function. The Fermi energy has been aligned to zero.}
\end{figure*}
To examine the validity of Eq. (\ref{eq-16}), we calculate the $\mathcal{T}$-odd spin conductivity for realistic altermagnets using both Eqs. (\ref{eq-14}) and (\ref{eq-16}). As an illustration, we consider the orthorhombic MnPd$_2$ ($^2m^2m^1m$)\cite{jpcs212}, tetragonal GdAlSi ($^24/^1m^2m^1m$)\cite{prb224436} and OsO$_2$ ($^24/^1m^2m^1m$)\cite{prm034407}, where the parentheses indicate the corresponding spin point groups\cite{prx031042,prx040501}. Those altermagnets are metallic and reveal sizable nonrelativistic spin splittings\cite{prb224436,prm034407}. For MnPd$_2$, the opposite-spin sublattices are connected by the spin-group operation $[C_2||C_{2y}t]$ with the translation $t=(0, 1/2, 1/2)$ [see Fig. \ref{f-2}(a)]. For GdAlSi, the opposite-spin sublattices are connected by the spin-group operation $[C_2||C_{4z}t]$ with $t=(0, 1/2, 1/4)$ [see Fig. \ref{f-2}(b)]. For OsO$_2$, the opposite-spin sublattices are connected by the spin-group operation $[C_2||C_{4z}t]$ with $t=(1/2, 1/2, 1/2)$ [see Fig. \ref{f-2}(c)].

Our density functional theory calculations were performed using the plane-wave ultrasoft pseudopotential method\cite{prb7892} as implemented in the QUANTUM ESPRESSO\cite{jpcm395502,jpcm465901,jcp154105}. An energy cutoff of $50$ Ry for the plane-wave expansion and generalized gradient approximation (GGA)\cite{prl3865} for the exchange and correlation functional were adopted throughout. For the orthorhombic MnPd$_2$\cite{jpcs212}, we used the lattice constants $a=8.123$, $b=5.481$ and $c=4.082$ {\AA} and the $k$-point grid of $5\times8\times10$. For the tetragonal GdAlSi\cite{prb224436}, we used the lattice constants $a=4.137$, $c=14.433$ {\AA} and the $k$-point grid of $14\times14\times4$. For the tetragonal OsO$_2$\cite{prm034407}, we used the lattice constants $a=4.565$, $c=3.211$ {\AA} and the $k$-point grid of $8\times8\times11$. A Hubbard-$U$ correction of $U_{eff}=7.0$ eV ($3.0$ eV) on the Gd-$4f$ (Os-$5d$) orbital was adopted to capture the electron correlation.

The spin conductivity was calculated using the Wannier-Linear-Response code\cite{prl187204,wtl} while the spin-resolved electrical conductivity was calculated by using the Boltzmann transport theory under the relaxation time approximation, as implemented in the BoltzWann module\cite{BoltzWann} of the Wannier90 code\cite{jpcm165902}. The tight-binding Hamiltonian was constructed from the maximally localized Wannier function basis set\cite{prb12847,rmp1419}. The $k$-point grids of $200\times320\times400$, $500\times500\times150$ and $280\times280\times400$ for evaluating the spin and electrical conductivities were used for MnPd$_2$, GdAlSi and OsO$_2$, respectively.

Since the magnetic easy axes for all the above altermagnets are along the $c$ axis ($z$ direction), we have $a=z$ and $\alpha=0$ in Eq. (\ref{eq-16}). Moreover, the spin-group analysis suggests that the nonzero $\mathcal{T}$-odd spin conductivity component for the spin point groups $^2m^2m^1m$ and $^24/^1m^2m^1m$ is $\sigma_{xy}^{z, \text{odd}}$\cite{prb224423,apr031412}. Figure \ref{f-2}(d-f) show the calculated $\sigma_{xy}^{z, \text{odd}}$ without SOC as a function of energy for MnPd$_2$ [Fig. \ref{f-2}(a)], GdAlSi [Fig. \ref{f-2}(b)] and OsO$_2$ [Fig. \ref{f-2}(c)]. One finds rather good agreement between $\sigma_{xy}^{z, \text{odd}}$ calculated using Eq. (\ref{eq-14}) and that calculated using Eq. (\ref{eq-16}). This confirms the quantitative validity of Eq. (\ref{eq-16}) to calculate the $\mathcal{T}$-odd spin conductivity in the nonrelativistic limit. On the other hand, Eq. (\ref{eq-16}) not only facilitates calculation of the $\mathcal{T}$-odd spin conductivity in the nonrelativistic limit, but also provides a simple physical picture to understand the nonrelativistic charge-to-spin conversion mechanism in magnetic materials, particularly in altermagnets. It should be noticed that the effect of SOC on the spin conductivity is twofold: to change its magnitude, and to yield additional nonzero spin conductivity components. For the magnetic materials containing the heavy element, the effect on the magnitude of the spin conductivity is expected to be significant. Moreover, there are additional nonzero spin conductivity components by considering the SOC. In the relativistic case, the spin conductivity tensor should be dictated by the magnetic point group\cite{prl127701}.

\section{Conclusion\label{sec4}}
In summary, we present an alternative but simplified derivation of the spin conductivity expression based on the adiabatic perturbation theory. We show that the $\mathcal{T}$-even expression has exactly the same form as those derived from the Kubo linear response theory and quantum kinetic theory. The $\mathcal{T}$-odd expression contains only intraband contribution, which is distinct from that derived from the Kubo linear response theory. However, the $\mathcal{T}$-odd expression agrees exactly with previous ones in the clean limit. We further simplify the $\mathcal{T}$-odd expression in the nonrelativistic limit and confirm its quantitative validity based on DFT calculations. Our work provides an alternative way to gain some insight into the physics of the charge-to-spin conversion in magnetic materials.

\begin{center}
{\bf Acknowledgements}
\end{center}

This work was supported by the National Natural Science Foundation of China (No. 12674135 and No. 12274102). The atomic structures were produced using the VESTA software\cite{vesta}.

\begin{center}
{\bf Conflict of Interest}
\end{center}

The authors declare no conflict of interest.

\begin{center}
{\bf Data Availability Statement}
\end{center}

The data that support the findings of this article are not publicly available. The data are available from the authors upon reasonable request.

\begin{widetext}
\appendix
\section{Derivation of Eq. (\ref{eq-3}) \label{secA}}
We substitute the eigenstate $|\Psi(t)\rangle$ Eq. (\ref{eq-2}) into the time-dependent Schrödinger equation $i\hbar\partial_t|\Psi(t)\rangle=\mathcal{H}|\Psi(t)\rangle$. On the left side, we have
\begin{equation}\label{qq-1}
\begin{aligned}
  i\hbar\partial_t|\Psi(t)\rangle&=e^{i[\theta_{n\mathbf{k}}(t)+\gamma_{n\mathbf{k}}(t)]}[(\epsilon_{n\mathbf{k}}-\hbar\dot{k}_aA_{nk_a}+i\hbar\dot{k}_a\partial_{k_a})|\psi_{n\mathbf{k}}\rangle+(\epsilon_{n\mathbf{k}}\dot{k}_b+i\hbar\dot{k}_b\partial_t+i\hbar\ddot{k}_b-\hbar\dot{k}_a\dot{k}_bA_{nk_a})|\delta\psi_{nk_b}\rangle]\\
  &\approx e^{i[\theta_{n\mathbf{k}}(t)+\gamma_{n\mathbf{k}}(t)]}[(\epsilon_{n\mathbf{k}}-\hbar\dot{k}_aA_{nk_a}+i\hbar\dot{k}_a\partial_{k_a})|\psi_{n\mathbf{k}}\rangle+\epsilon_{n\mathbf{k}}\dot{k}_a|\delta\psi_{nk_a}\rangle],
\end{aligned}
\end{equation}
where the second-order terms proportional to $\dot{k}_a\dot{k}_b$, $\ddot{k}_a$ and $\dot{k}_a\partial_t|\delta\psi_{nk_a}\rangle$ are disregarded and $\ddot{k}_a$ indicates a second derivative with respect to $t$. The right side of the Schrödinger equation gives rise to
\begin{equation}\label{aq-2}
\mathcal{H}|\Psi(t)\rangle=e^{i[\theta_{n\mathbf{k}}(t)+\gamma_{n\mathbf{k}}(t)]}[\epsilon_{n\mathbf{k}}|\psi_{n\mathbf{k}}\rangle+\dot{k}_a\mathcal{H}|\delta\psi_{nk_a}\rangle].
\end{equation}
From $i\hbar\partial_t|\Psi(t)\rangle=\mathcal{H}|\Psi(t)\rangle$ and removing the common factors, we obtain
\begin{equation}\label{aq-3}
  (\epsilon_{n\mathbf{k}}-\mathcal{H})|\delta\psi_{nk_a}\rangle=-i\hbar(\partial_{k_a}+iA_{nk_a})|\psi_{n\mathbf{k}}\rangle.
\end{equation}

\section{Derivation of Eq. (\ref{eq-4}) \label{secB}}
We insert the completeness relation $\sum_{m\mathbf{k'}}|\psi_{m\mathbf{k'}}\rangle\langle\psi_{m\mathbf{k'}}|=1$ on both sides of Eq. (\ref{eq-3})
\begin{equation}\label{bq-1}
  \sum_{m\mathbf{k'}}|\psi_{m\mathbf{k'}}\rangle\langle\psi_{m\mathbf{k'}}|(\epsilon_{n\mathbf{k}}-\mathcal{H})|\delta\psi_{nk_a}\rangle=-i\hbar\sum_{m\mathbf{k'}}|\psi_{m\mathbf{k'}}\rangle\langle\psi_{m\mathbf{k'}}|(\partial_{k_a}+iA_{nk_a})|\psi_{n\mathbf{k}}\rangle.
\end{equation}
Thus, we have
\begin{equation}\label{bq-2}
  (\epsilon_{n\mathbf{k}}-\epsilon_{m\mathbf{k'}})|\delta\psi_{nk_a}\rangle=-i\hbar\sum_{m\mathbf{k'}}|\psi_{m\mathbf{k'}}\rangle\langle\psi_{m\mathbf{k'}}|(\partial_{k_a}+iA_{nk_a})|\psi_{n\mathbf{k}}\rangle.
\end{equation}
We now use the orthonormality relations $\langle\psi_{m\mathbf{k'}}|\partial_{k_a}|\psi_{n\mathbf{k}}\rangle=-i\langle\psi_{m\mathbf{k'}}|r_a|\psi_{n\mathbf{k}}\rangle=N\delta_{\mathbf{kk'}}\langle u_{m\mathbf{k}}|\partial_{k_a}|u_{n\mathbf{k}}\rangle$ (for proof, see the following) and $\langle\psi_{m\mathbf{k'}}|iA_{nk_a}|\psi_{n\mathbf{k}}\rangle\propto\delta_{mn}\delta_{\mathbf{kk'}}$ to reduce this further to
\begin{equation}\label{bq-3}
  |\delta\psi_{nk_a}\rangle=-i\hbar\sum_{m\neq n}\frac{|\psi_{m\mathbf{k}}\rangle\langle\psi_{m\mathbf{k}}|\partial_{k_a}|\psi_{n\mathbf{k}}\rangle}{\epsilon_{n\mathbf{k}}-\epsilon_{m\mathbf{k}}}.
\end{equation}
{\bf Proof of $\langle\psi_{m\mathbf{k'}}|r_a|\psi_{n\mathbf{k}}\rangle=N\delta_{\mathbf{kk'}}\langle u_{m\mathbf{k}}|i\partial_{k_a}|u_{n\mathbf{k}}\rangle$}. The left side turns out to be
\begin{equation}\label{bq-4}
\begin{aligned}
 \langle\psi_{m\mathbf{k'}}|r_a|\psi_{n\mathbf{k}}\rangle&=\int_{N\Omega}\psi^*_{m\mathbf{k'}}(\mathbf{r})r_a\psi_{n\mathbf{k}}(\mathbf{r})d^3r=\int_{N\Omega}u^*_{m\mathbf{k'}}(\mathbf{r})e^{-i\mathbf{k'\cdot r}}r_ae^{i\mathbf{k\cdot r}}u_{n\mathbf{k}}(\mathbf{r})d^3r\\
 &=\int_{N\Omega}u^*_{m\mathbf{k'}}(\mathbf{r})e^{-i\mathbf{k'\cdot r}}(-i\partial_{k_a}e^{i\mathbf{k\cdot r}})u_{n\mathbf{k}}(\mathbf{r})d^3r\\
 &=-i\partial_{k_a}\int_{N\Omega}\psi^*_{m\mathbf{k'}}(\mathbf{r})\psi_{n\mathbf{k}}(\mathbf{r})d^3r+\int_{N\Omega}e^{i\mathbf{(k-k')\cdot r}}u^*_{m\mathbf{k'}}(\mathbf{r})i\partial_{k_a}u_{n\mathbf{k}}(\mathbf{r})d^3r\\
 &=\sum_\mathbf{R}e^{i\mathbf{(k-k')\cdot R}}\int_{\Omega}e^{i\mathbf{(k-k')\cdot r}}u^*_{m\mathbf{k'}}(\mathbf{r})i\partial_{k_a}u_{n\mathbf{k}}(\mathbf{r})d^3r\\
 &=N\delta_{\mathbf{kk'}}\langle u_{m\mathbf{k}}|i\partial_{k_a}|u_{n\mathbf{k}}\rangle,\\
\end{aligned}
\end{equation}
where $\Omega$ is the volume of the unit cell.

\section{Derivation of Eq. (\ref{eq-8}) \label{secC}}
From the orthonormality relation $\langle\psi_{m\mathbf{k}}|\psi_{n\mathbf{k}}\rangle=\delta_{mn}$, we have $\partial_{k_a}\langle\psi_{m\mathbf{k}}|\psi_{n\mathbf{k}}\rangle=\langle\partial_{k_a}\psi_{m\mathbf{k}}|\psi_{n\mathbf{k}}\rangle+\langle\psi_{m\mathbf{k}}|\partial_{k_a}\psi_{n\mathbf{k}}\rangle=0$. This yields $\langle\partial_{k_a}\psi_{m\mathbf{k}}|\psi_{n\mathbf{k}}\rangle=-\langle\psi_{m\mathbf{k}}|\partial_{k_a}\psi_{n\mathbf{k}}\rangle$. Moreover, from the identity $\langle\psi_{m\mathbf{k}}|\mathcal{H}|\psi_{n\mathbf{k}}\rangle=\epsilon_{n\mathbf{k}}\delta_{mn}$, it is apparent that $\partial_{k_a}\langle\psi_{m\mathbf{k}}|\mathcal{H}|\psi_{n\mathbf{k}}\rangle=0$. Then, we have
\begin{equation}\label{cq-1}
\begin{aligned}
 \partial_{k_a}\langle\psi_{m\mathbf{k}}|\mathcal{H}|\psi_{n\mathbf{k}}\rangle&=\langle\partial_{k_a}\psi_{m\mathbf{k}}|\mathcal{H}|\psi_{n\mathbf{k}}\rangle+\langle\psi_{m\mathbf{k}}|\partial_{k_a}\mathcal{H}|\psi_{n\mathbf{k}}\rangle+\langle\psi_{m\mathbf{k}}|\mathcal{H}|\partial_{k_a}\psi_{n\mathbf{k}}\rangle\\
 &=\epsilon_{n\mathbf{k}}\langle\partial_{k_a}\psi_{m\mathbf{k}}|\psi_{n\mathbf{k}}\rangle+\epsilon_{m\mathbf{k}}\langle\psi_{m\mathbf{k}}|\partial_{k_a}\psi_{n\mathbf{k}}\rangle+\langle\psi_{m\mathbf{k}}|\partial_{k_a}\mathcal{H}|\psi_{n\mathbf{k}}\rangle\\
 &=(\epsilon_{m\mathbf{k}}-\epsilon_{n\mathbf{k}})\langle\psi_{m\mathbf{k}}|\partial_{k_a}\psi_{n\mathbf{k}}\rangle+\langle\psi_{m\mathbf{k}}|\partial_{k_a}\mathcal{H}|\psi_{n\mathbf{k}}\rangle=0.
\end{aligned}
\end{equation}
Hence,
\begin{equation}\label{cq-2}
\langle\psi_{m\mathbf{k}}|\partial_{k_a}\psi_{n\mathbf{k}}\rangle=\frac{\langle\psi_{m\mathbf{k}}|\partial_{k_a}\mathcal{H}|\psi_{n\mathbf{k}}\rangle}{\epsilon_{n\mathbf{k}}-\epsilon_{m\mathbf{k}}}.
\end{equation}
\end{widetext}


\begin{thebibliography}{99}
\bibitem{rmp1213} J. Sinova, S. O. Valenzuela, J. Wunderlich, C. H. Back, and T. Jungwirth, Spin Hall effects, \emph{Rev. Mod. Phys.} {\bf 2015}, 87, 1213.
\bibitem{npj27} W. Han, Y. Otani, and S. Maekawa, Quantum materials for spin and charge conversion, \emph{npj Quant. Mater.} {\bf 2018}, 3, 27.
\bibitem{nrm258} F. Trier, P. Noël, J. Kim, J. Attané, L. Vila, and M. Bibes, Oxide spin-orbitronics: spin –charge interconversion and topological spin textures, \emph{Nat. Rev. Mater.} {\bf 2022}, 7, 258.
\bibitem{jpcm179} R. H. Silsbee, Spin-orbit induced coupling of charge current and spin polarization, \emph{J. Phys.: Condens. Matter} {\bf 2004}, 16, R179.
\bibitem{prb085438} L. L. Tao and Evgeny Y. Tsymbal, Spin-orbit dependence of anisotropic current-induced spin polarization, \emph{Phys. Rev. B} {\bf 2021}, 104, 085438.
\bibitem{prb212405} A. Manchon and S. Zhang, Theory of nonequilibrium intrinsic spin torque in a single nanomagnet, \emph{Phys. Rev. B} {\bf 2008}, 78, 212405.
\bibitem{prb094422} A. Manchon and S. Zhang, Theory of spin torque due to spin-orbit coupling, \emph{Phys. Rev. B} {\bf 2009}, 79, 094422.
\bibitem{prb094424} A. Matos-Abiague and R. L. Rodr\'{\i}guez-Su\'arez, Spin-orbit coupling mediated spin torque in a single ferromagnetic layer, \emph{Phys. Rev. B} {\bf 2009}, 80, 094424.
\bibitem{nature189} I. M. Miron, K. Garello, G. Gaudin, P. J. Zermatten, M. V. Costache, S. Auffret, S. Bandiera, B. Rodmacq, A. Schuhl, and P. Gambardella, Perpendicular switching of a single ferromagnetic layer induced by in-plane current injection, \emph{Nature} {\bf 2011}, 476, 189.
\bibitem{science555} L. Liu, C. F. Pai, Y. Li, H. W. Tseng, D. C. Ralph, and R. A. Buhrman, Spin-torque switching with the giant spin Hall effect of tantalum, \emph{Science} {\bf 2012}, 336, 555.
\bibitem{prl096602} L. Liu, O. J. Lee, T. J. Gudmundsen, D. C. Ralph, and R. A. Buhrman, Current-Induced Switching of Perpendicularly Magnetized Magnetic Layers Using Spin Torque from the Spin Hall Effect, \emph{Phys. Rev. Lett.} {\bf 2012}, 109, 096602.
\bibitem{rmp035004} A. Manchon, J. \ifmmode \check{Z}\else \v{Z}\fi{}elezn\'y, I. M. Miron, T. Jungwirth, J. Sinova, A. Thiaville, K. Garello, and P. Gambardella, Current-induced spin-orbit torques in ferromagnetic and antiferromagnetic systems, Rev. Mod. Phys. {\bf 2019}, 91, 035004.
\bibitem{prb144411} Y.-T. Chen, S. Takahashi, H. Nakayama, M. Althammer, S. T. B. Goennenwein, E. Saitoh, and G. E. W. Bauer, Theory of spin Hall magnetoresistance, \emph{Phys. Rev. B} {\bf 2013}, 87, 144411.
\bibitem{np570} C. O. Avci, K. Garello, A. Ghosh, M. Gabureac, S. F. Alvarado, and P. Gambardella, Unidirectional spin Hall magnetoresistance in ferromagnet/normal metal bilayers, \emph{Nat. Phys.} {\bf 2015}, 11, 570.
\bibitem{apl120502} X. Han, X. Wang, C. Wang, G. Yu, X. Lv, Spin-orbit torques: Materials, physics, and devices, \emph{Appl. Phys. Lett.} {\bf 2021}, 118, 120502.
\bibitem{pms100761} C. Song, R. Zhang, L. Liao, Y. Zhou, X. Zhou, R. Chen, Y. You, X. Chen, and F. Pan, Spin-orbit torques: Materials, mechanisms, performances, and potential applications, \emph{Prog. Mater. Sci.} {\bf 2021}, 118, 100761.
\bibitem{npj8} M.-G. Kang, S. Lee, and B.-G. Park, Field-free spin-orbit torques switching and its applications, \emph{npj Spintronics} {\bf 2025}, 3, 8.
\bibitem{apr011305} Z. Han, Y. Huo, Y. Yu, H. Wu, L. Gao, Z. Ma, G. Zhang, M. Jiang, Unveiling the potential of spin–orbit torque in a magnetic single layer for advancing spintronics application, \emph{Appl. Phys. Rev.} {\bf 2026}, 13, 011305.
\bibitem{prb024410} L. Salemi and P. M. Oppeneer, Theory of magnetic spin and orbital Hall and Nernst effects in bulk ferromagnets, \emph{Phys. Rev. B} {\bf 2022}, 106, 024410.
\bibitem{prb174423} F. Freimuth, S. Blügel, and Y. Mokrousov, Spin-orbit torques in Co/Pt(111) and Mn/W(001) magnetic bilayers from first principles, \emph{Phys. Rev. B} {\bf 2014}, 90, 174423.
\bibitem{prb134402} H. Li, H. Gao, L. P. Z\^arbo, K. V\'yborn\'y, X. Wang, I. Garate, F. Do\ifmmode \check{g}\else \v{g}\fi{}an, A. \ifmmode \check{C}\else \v{C}\fi{}ejchan, J. Sinova, T. Jungwirth, and A. Manchon, Intraband and interband spin-orbit torques in noncentrosymmetric ferromagnets, \emph{Phys. Rev. B} {\bf 2015}, 91, 134402.
\bibitem{prl187204} J. \ifmmode \check{Z}\else \v{Z}\fi{}elezn\'y, Y. Zhang, C. Felser, and B. Yan, Spin-Polarized Current in Noncollinear Antiferromagnets, \emph{Phys. Rev. Lett.} {\bf 2017}, 119, 187204.
\bibitem{na627} M. Kimata, H. Chen, K. Kondou, S. Sugimoto, P. K. Muduli, M. Ikhlas, Y. Omori, T. Tomita, Allan. H. MacDonald, S. Nakatsuji, and Y. Otani, Magnetic and magnetic inverse spin Hall effects in a non-collinear antiferromagnet, \emph{Nature} {\bf 2019}, 565, 627.
\bibitem{prl127701} R. Gonz\'alez-Hern\'andez, L. \ifmmode \check{S}\else \v{S}\fi{}mejkal, K. V\'yborn\'y, Y. Yahagi, J. Sinova, T. Jungwirth, and J. \ifmmode \check{Z}\else \v{Z}\fi{}elezn\'y, Efficient Electrical Spin Splitter Based on Nonrelativistic Collinear Antiferromagnetism, \emph{Phys. Rev. Lett.} {\bf 2021}, 126, 127701.
\bibitem{prb224401} F. Zheng, M. Zhu, J. Dong, X. Li, Y. Zhou, K. Wu, and J. Zhang, Anatomy of the spin Hall effect in ferromagnetic metals, \emph{Phys. Rev. B} {\bf 2024}, 109, 224401.
\bibitem{npj46} K. Tenzin, B. Kilic, R. M. Sattigeri, Z. He, C. C. Ye, M. Costa, M. B. Nardelli, C. Autieri, and J. Sławińska, Persistent spin textures, altermagnetism and charge-to-spin conversion in metallic chiral crystals TM$_3$X$_6$, \emph{npj Spintronics} {\bf 2025}, 3, 46.
\bibitem{prx031042} L. \ifmmode \check{S}\else \v{S}\fi{}mejkal, J. Sinova, and T. Jungwirth, Beyond Conventional Ferromagnetism and Antiferromagnetism: A Phase with Nonrelativistic Spin and Crystal Rotation Symmetry, \emph{Phys. Rev. X} {\bf 2022}, 12, 031042.
\bibitem{prx040501} L. \ifmmode \check{S}\else \v{S}\fi{}mejkal, J. Sinova, and T. Jungwirth, Emerging Research Landscape of Altermagnetism, \emph{Phys. Rev. X} {\bf 2022}, 12, 040501.
\bibitem{afm2409327} L. Bai, W. Feng, S. Liu, L. \ifmmode \check{S}\else \v{S}\fi{}mejkal, Y. Mokrousov, Y. Yao, Altermagnetism: Exploring New Frontiers in Magnetism and Spintronics, \emph{Adv. Funct. Mater.} {\bf 2024}, 34, 2409327.
\bibitem{jcas2257} S. S. Fender, O. Gonzalez, and D. K. Bediako, Altermagnetism: A Chemical Perspective, \emph{J. Am. Chem. Soc.} {\bf 2025}, 147, 2257.
\bibitem{nrm2025} C. Song, H. Bai, Z. Zhou, L. Han, H. Reichlova, J. Hugo Dil, J. Liu, X. Chen, and F. Pan, Altermagnets as a new class of functional materials, \emph{Nat. Rev. Mater.} {\bf 2025}, 10, 473.
\bibitem{nature837} T. Jungwirth, J. Sinova, R. M. Fernandes, Q. Liu, H. Watanabe, S. Murakami, S. Nakatsuji, and L. Šmejkal, Symmetry, microscopy and spectroscopy signatures of altermagnetism, \emph{Nature} {\bf 2026}, 649, 837.
\bibitem{jpsj123702} S. Hayami, Y. Yanagi, and H. Kusunose, Momentum-Dependent Spin Splitting by Collinear Antiferromagnetic Ordering, \emph{J. Phys. Soc. Jpn.} {\bf 2019}, 88, 123702.
\bibitem{prb014422} L.-D. Yuan, Z. Wang, J.-W. Luo, E. I. Rashba, and A. Zunger, Giant momentum-dependent spin splitting in centrosymmetric low-Z antiferromagnets, \emph{Phys. Rev. B} {\bf 2020}, 102, 014422.
\bibitem{nc2846} H.-Y. Ma, M. Hu, N. Li, J. Liu, W. Yao, J.-F. Jia, and J. Liu, Multifunctional antiferromagnetic materials with giant piezomagnetism and noncollinear spin current, \emph{Nat. Commun.} {\bf 2021}, 12, 2846.
\bibitem{prl197202} H. Bai, L. Han, X. Y. Feng, Y. J. Zhou, R. X. Su, Q. Wang, L. Y. Liao, W. X. Zhu, X. Z. Chen, F. Pan, X. L. Fan, and C. Song, Observation of Spin Splitting Torque in a Collinear Antiferromagnet ${\mathrm{RuO}}_{2}$, \emph{Phys. Rev. Lett.} {\bf 2022}, 128, 197202.
\bibitem{prl137201} S. Karube, T. Tanaka, D. Sugawara, N. Kadoguchi, M. Kohda, and J. Nitta, Observation of Spin-Splitter Torque in Collinear Antiferromagnetic ${\mathrm{RuO}}_{2}$, \emph{Phys. Rev. Lett.} {\bf 2022}, 129, 137201.
\bibitem{nc5646} Y. Zhang, H. Bai, J. Dai, L. Han, C. Chen, S. Liang, Y. Cao, Y. Zhang, Q. Wang, W. Zhu, F. Pan, and C. Song, Electrical manipulation of spin splitting torque in altermagnetic RuO$_2$, \emph{Nat. Commun.} {\bf 2025}, 16, 5646.
\bibitem{nc1309} Y. Guo, A. Chen, Z. Zeng, T. An, Q. Cui, Y. Ga, D. Zheng, X. Deng, X. Zhang, M. Tang, Z. Zhu, C. Wu, J. Zhang, Y. Fan, Z. Wang, W. Song, P. Yan, T. Zhu, S. Wang, X. Han, J. Zhao, Kang L. Wang, G. Yu, X. Zhang, and H. Wu, Magnetic memory driven by spin splitting torque in nonrelativistic collinear antiferromagnet, \emph{Nat. Commun.} {\bf 2026}, 17, 1309.
\bibitem{prb014416} A. Cr\'epieux and P. Bruno, Theory of the anomalous Hall effect from the Kubo formula and the Dirac equation, \emph{Phys. Rev. B} {\bf 2011}, 64, 014416.
\bibitem{Vanderbilt} D. Vanderbilt, \emph{Berry Phases in Electronic Structure Theory: Electric Polarization, Orbital Magnetization and Topologi cal Insulators} (Cambridge University Press, Cambridge, UK, 2018).
\bibitem{ase14818} X.-B. Qiang, T. Liu, Z.-X. Gao, H.-Z. Lu, and X. C. Xie, A Clarification on Quantum-Metric-Induced Nonlinear Transport, \emph{Adv. Sci.} {\bf 2026}, 13, e14818.
\bibitem{Berry} M. V. Berry, Quantal phase factors accompanying adiabatic changes, \emph{Proc. R. Soc. Lond. A} {\bf 1984}, 392, 45.
\bibitem{rmd1959} D. Xiao, M.-C. Chang, and Q. Niu, Berry phase effects on electronic properties, \emph{Rev. Mod. Phys.} {\bf 2010}, 82, 1959.
\bibitem{prb224423} M. Dou, X. Wang, and L. L. Tao, Anisotropic spin-polarized conductivity in collinear altermagnets, \emph{Phys. Rev. B} {\bf 2025}, 111, 224423.
\bibitem{apr031412} M. Dou, X. Chen, Z. Liu, Y. Sui, X. Wang, and L. L. Tao, Highly anisotropic nonrelativistic charge-to-spin conversion in altermagnets, \emph{Appl. Phys. Rev.} {\bf 2026}, 13, 031412.
\bibitem{callaway} J. Callaway, \emph{Quantum Theory of the Solid State} (Academic, San Diego, 1991).
\bibitem{prb155411} X. Chen, Y. Fu, X. Wang, Yu Sui, H. J. Zhao, and L. L. Tao, Magnetic control of nonreciprocal charge transport, \emph{Phys. Rev. B} {\bf 2025}, 111, 155411.
\bibitem{jpcs212} G. Kádár, E. Krén, and M. Márton, New antiferromagnetic intermetallic compound in the Mn-Pd system: MnPd$_2$, \emph{J. Phys. Chem. Solids} {\bf 1972}, 33, 212.
\bibitem{prb224436} J. Nag, B. Das, S. Bhowal, Y. Nishioka, B. Bandyopadhyay, S. Sarker, S. Kumar, K. Kuroda, V. Gopalan, A. Kimura, K. G. Suresh, and A. Alam, GdAlSi: An antiferromagnetic topological Weyl semimetal with nonrelativistic spin splitting, \emph{Phys. Rev. B} {\bf 2024}, 110, 224436.
\bibitem{prm034407} P. R. Raghuvanshi, T. Berlijn, D. S. Parker, S. Wang, M. E. Manley, R. P. Hermann, L. Lindsay, and V. R. Cooper, Altermagnetic behavior in ${\mathrm{OsO}}_{2}$: Parallels with ${\mathrm{RuO}}_{2}$, \emph{Phys. Rev. Materials} {\bf 2025}, 9, 034407.
\bibitem{prb7892} D. Vanderbilt, Soft self-consistent pseudopotentials in a generalized eigenvalue formalism, \emph{Phys. Rev. B} {\bf 1990}, 41, 7892(R).
\bibitem{jpcm395502} P. Giannozzi, S. Baroni, N. Bonini, M. Calandra, R. Car, C. Cavazzoni, D. Ceresoli, G. L. Chiarotti, M. Cococcioni, I. Dabo \emph{et al.}, QUANTUM ESPRESSO: a modular and open-source software project for quantum simulations of materials, \emph{J. Phys.: Condens. Matter} {\bf 2009}, 21, 395502.
\bibitem{jpcm465901} P. Giannozzi, O. Andreussi, T. Brumme, O. Bunau, M. B. Nardelli, M. Calandra, R. Car, C. Cavazzoni, D. Ceresoli, M. Cococcioni \emph{et al.}, Advanced capabilities for materials modelling with Quantum ESPRESSO, \emph{J. Phys.: Condens. Matter} {\bf 2017}, 29, 465901.
\bibitem{jcp154105} P. Giannozzi, O. Baseggio, P. Bonf\`a, D. Brunato, R. Car, I. Carnimeo, C. Cavazzoni, S. de Gironcoli, P. Delugas, F. F. Ruffino \emph{et al.}, Quantum ESPRESSO toward the exascale, \emph{J. Chem. Phys.} {\bf 2020}, 152 154105.
\bibitem{prl3865} J. P. Perdew, K. Burke, and M. Ernzerhof, Generalized Gradient Approximation Made Simple, \emph{Phys. Rev. Lett.} {\bf 1996}, 77, 3865.
\bibitem{wtl} J. Zelezny, see https://bitbucket.org/zeleznyj/wannier-linearresponse/wiki/Home for the Wannier-linear-response code.
\bibitem{BoltzWann} G. Pizzi, D. Volja, B. Kozinsky, M. Fornari, and N. Marzari, BoltzWann: A code for the evaluation of thermoelectric and electronic transport properties with a maximally-localized Wannier functions basis, \emph{Comput. Phys. Commun.} {\bf 2014}, 185, 422.
\bibitem{jpcm165902} P. Giovanni, V. Vitale, R. Arita, S. Blügel, F. Freimuth, G. Granton, M. Gibertini, D. Gresch, C. Johnson, T. Koretsune \emph{et al.}, Wannier90 as a community code: new features and applications, \emph{J. Phys.: Condens. Matter} {\bf 2020}, 32, 165902.
\bibitem{prb12847} N. Marzari and D. Vanderbilt, Maximally localized generalized Wannier functions for composite energy bands, \emph{Phys. Rev. B} {\bf 1997}, 56, 12847.
\bibitem{rmp1419} N. Marzari, A. A. Mostofi, J. R. Yates, I. Souza and D. Vanderbilt, Maximally localized Wannier functions: Theory and applications, \emph{Rev. Mod. Phys.} {\bf 2012}, 84, 1419.
\bibitem{vesta} K. Momma and F. Izumi, VESTA 3 for three-dimensional visualization of crystal, volumetric and morphology data, \emph{J. Appl. Crystallogr.} {\bf 2011}, 44, 1272.
\end{thebibliography}
\end{document}